\documentclass[twocolumn]{aastex6}
\shorttitle{Cluster-Forming Clump in S235}
\shortauthors{Shimoikura et al.}

\begin{document}


\title{Discovery of Infalling Motion with Rotation of the Cluster-Forming Clump S235AB and Its Implication to the Clump Structures}

\author{
Tomomi Shimoikura\altaffilmark{1}, Kazuhito Dobashi\altaffilmark{1}, 
Tomoaki Matsumoto\altaffilmark{2}, 
\and Fumitaka Nakamura\altaffilmark{3}
}

\affil{\scriptsize{\rm $^1$ ikura@u-gakugei.ac.jp}}
\altaffiltext{1}{Department of Astronomy and Earth Sciences, Tokyo Gakugei University, Koganei, Tokyo  184-8501, Japan}
\altaffiltext{2}{Department of Humanity and Environment, Hosei University, Fujimi, Chiyoda-ku, Tokyo 102-8160, Japan}
\altaffiltext{3}{National Astronomical Observatory of Japan, Mitaka, Tokyo 181-8588, Japan}


%%%%%%%%%%%%%%%%%%%%%%%%%%%%%%%%%%%
%                                                      Abstract                                 %
%%%%%%%%%%%%%%%%%%%%%%%%%%%%%%%%%%%

\begin{abstract}
We report the discovery of infalling motion with rotation of S235AB the massive cluster-forming clump
($\sim 1\times 10^3$ $M_\sun$) in the S235 region.
Our C$^{18}$O observations with the
45m telescope at the Nobeyama Radio Observatory have revealed
the elliptical shape of the clump. Position-velocity (PV) diagram taken along its
major axis exhibits two
well-defined peaks symmetrically located with respect to the clump center,
which is similar to that found for a dynamically infalling envelope
with rotation around a single protostar modeled by N. Ohashi and his collaborators,
indicating that the cluster-forming clump is also collapsing by the self-gravity
toward the clump center.
%as the envelopes of the single protostars
%inspite of the large difference in mass and size.
%Because the feature in the PV diagram
%is often observed in massive cluster-forming clumps, we suggest that the infalling motion with rotation
%is a common process in an early stage of cluster formation.
With analogue to Ohashi 's model,
we made a simple model of an infalling, rotating clump  to fit the observed data.
Based on the inferred model parameters as well as results of earlier observations and
simulations in the literature,
% provide us with important information on
%the structures of the clump. In this paper,
we discuss structures of the clump such as the relation among
the global mass infall rate ($\sim 1 \times 10^{-3}$ $M_\sun$ yr$^{-1}$),
formation of a compact core (with a mass and size of $\sim4$ $M_\sun$ and $\lesssim 0.1$ pc)  at the center,
and a massive star ($\sim 11$ $M_\sun$) forming in the core.
\end{abstract}


%%%%%%%%%%%%%%%%%%%%%%%%%%%%%%%%%%%
%                                         INTRODUCTION                                  %
%%%%%%%%%%%%%%%%%%%%%%%%%%%%%%%%%%%

\section{INTRODUCTION}\label{sec:introduction}
Dynamics of cluster-forming clumps is of particular interest,
because most of stars are formed in clusters \citep[e.g.,][]{Lada2003}.
Velocity field of massive cluster-forming clumps ($\sim1000$ $M_\sun$)
traced by some high density tracers
(e.g., H$^{13}$CO$^+$, C$^{18}$O, etc.)
is often complex,
exhibiting two or more velocity components.
These components are sometimes interpreted as
an evidence of clump-clump collisions \citep[e.g.,][]{Higuchi2009,Torii2011}, 
collisions of filaments \citep[e.g.,][]{Nakamura2014,Dobashi2014,Matsumoto2015},
or clumps blown by expanding compact {H{$\,${\sc ii}}} regions \citep[e.g.,][]{Shimoikura2015}.
It can be also possible that the velocity components are due to vibration of the clumps as suggested
for the small starless core B68 \citep{Lada2003b}.
Although all of these can be reasonable interpretations for some particular cases,
we should also consider the most likely possibility that they
are tracing the systematic internal motions of the clumps dominated by the self-gravity,
i.e., the dynamically infalling motions with rotation.

%or due to large-scale collapsing motions as found for a cluster-forming clump in NGC2264C \citep{Peretto2006,Peretto2007},
%which can be a reasonable interpretation for some particular cases.

%On-going cluster formations are often been observed toward clumps with a mass of $\sim1000$ $M_\sun$
%\citep[e.g.][]{Shimoikura2013}, and their dynamical motions are of particular interest because
%most of stars are formed in clusters \citep[e.g.,][]{Lada2003}.
In an early stage of cluster formation when the dynamics of the natal clump is not yet strongly influenced by
the internal cluster, 
%we would expect that a clump may undergo global gravitational contraction with rotation
%toward the center following the angular momentum conservation.
we would expect that the motion of the clump must follow simple kinetic laws
dominated by the self-gravity with increasing infall and rotation velocities toward the
center of the clump for conservation of the energy and angular momentum.
The high concentration of mass by the infall should form a central core
which will produce the most massive star in the cluster, and
the infalling mass of the natal clump should be once trapped by the core
and then transferred to the central star.

Such a scenario of the cluster-forming clumps can be supported by recent
numerical simulations \citep[e.g.,][]{Smith2009,Wang2010}.
For example, according to the simulations performed by \citet{Wang2010} who calculated
the evolution of a cluster-forming clump with a mass of $1.6 \times 10^3$ $M_\sun$,
%ここを要検討
a massive star can be naturally formed at the center of the clump.
Interestingly, they found that the mass of the massive star is controlled by the mass of the large-scale
natal clump rather than by the compact core defined as a 0.1-sized region around the star,
and the mass accretion rate onto the star is regulated
by the feedback of an outflow generated by the star.

We believe that the above picture on the structures of cluster-forming clumps
should be plausible. Because the infall and rotation velocities can be very large
around the center of the clumps, 
we should be able to detect large-scale systematic infalling motion
with rotation toward the center of cluster-forming clumps where the most massive star in the cluster is forming.
Such a detection should provide a good support to the qualitative picture of massive cluster-forming clumps,
but it has to be justified by the observations.

%If the findings by the simulations are ubiquitous,
%we should frequently detect large-scale systematic infalling motion 
%toward the center of cluster-forming clumps where the most massive star in the cluster should be forming.
%In addition, we would expect that the natal clumps should be rotating very fast
%around the clump center for the conservation of the angular momentum, which should also be observed.

The purpose of this paper is to report the detection of such an infalling motion with rotation of
a cluster-forming clump in the S235 region \citep{Sharpless}, and
to discuss its velocity and density structures
by fitting the observed data with a simple model.
% by comparing with the simulations of \citet{Wang2010}.
In our recent survey for cluster-forming clumps using
the 45m telescope at the Nobeyama Radio Observatory (NRO), 
we found that some cluster-forming clumps exhibit mainly two velocity components
(Shimoikura et al. 2016, in preparation),
and we realized that the spatial distributions of these
components are very similar to those found toward infalling envelopes with rotation
surrounding isolated single protostars \citep[e.g.,][]{Ohashi1997, Momose1998, Bernard1999},
indicating that the cluster-forming clumps should also be collapsing toward the clump center by the self-gravity
like in the case of the single protostars.
For these years, infalling motions of cluster-forming clumps have been discovered
by some authors \citep[e.g.,][]{Peretto2006, Peretto2007, Barnes2010, Reiter2011}, but their data 
would be more naturally understood if we consider the rotation in addition to the infall
as we found in this paper.

The clump studied here is one of the most typical cases
showing infalling motion with rotaion. It is cataloged as No.4423 \citep{Dobashi2011} or
TGU H1192P1 \citep{Dobashi2005} in the catalogs of dark clouds, and
two compact {H{$\,${\sc ii}}} regions called S235A and S235B \citep[e.g.,][]{Felli1997, Klein2005, saito2007}
are located in the clump (see Figure \ref{fig:fig1}).
We call this clump S235AB in this paper.
An embedded star cluster has been identified \citep[e.g.,][]{Camargo2011,Dewangan2011,Chavarra2014},
and the most massive member of the cluster named S235AB-MIR \citep{Felli2006} is located at the clump center
\citep[$\sim11$ $M_\sun$, ][]{Dewangan2011}.
Several signposts of early stellar evolution such as a molecular outflow
\citep[e.g.,][]{Nakano1986,Felli2004} and
a fast jet traced by water maser emission \citep[$\sim50$ km s$^{-1}$, ][]{Burns2015} have also been found.
S235AB should therefore be suited to detect and quantify the infalling motion with rotation.
% and to compare with
%the results of the simulations by \citet{Wang2010}.
Estimates of the distance to the clump vary in the range 1.6 -- 2.5 kpc \citep[e.g.,][]{Israel1978, Burns2015},
and we adopt a value of 1.8 kpc in this paper following \cite{Evans1981}.

%%%%%%%%%%%%%%%%%%%%%%%%%%%%%%%%%%%
%                              OBSERVATIONS                                                            %
%%%%%%%%%%%%%%%%%%%%%%%%%%%%%%%%%%%

\section{OBSERVATIONS} \label{sec:observations}
%% In this section, we use  the \subsection command to set off
%% a subsection.  \footnote is used to insert a footnote to the text.

In order to carry out a statistical study of cluster-forming clumps found in
a catalog of dense cores \citep{Dobashi2011,Dobashi2013},
we made molecular line observations of several massive clumps
including S235AB with the NRO 45m telescope
for 10 days in 2013 February. 
We observed the clumps in some molecular lines including $^{12}$CO$(J=1-0)$, 
$^{13}$CO$(J=1-0)$, and C$^{18}$O$(J=1-0)$
with a Superconductor-Insulator-Superconductor (SIS)
receiver named TZ \citep{Asayama}.
We used spectrometers called SAM45 which have 4096 channels and a frequency resolution of 7.6 kHz.
We performed a standard On-The-Fly mapping \citep{Sawada} around the target clumps,
and convolved the data with a spheroidal function to resample
them at the $7 \farcs 5$ grid along the equatorial coordinates.

More details of the observations will be given in a subsequent publication
together with results of statistics of the observed clumps (Shimoikura et al. 2016, in preparation).
In this paper, we will concentrate on the analyses of the velocity field of
S235AB mainly using the C$^{18}$O data.
The angular resolution and the noise levels of the C$^{18}$O data
are 22\arcsec and $\Delta T_{\rm mb}\simeq$ 0.3 K at a velocity resolution of 0.1 km s$^{-1}$.

%%%%%%%%%%%%%%%%%%%%%%%%%%%%%%%%%%%
%                                         RESULTS                                                 %
%%%%%%%%%%%%%%%%%%%%%%%%%%%%%%%%%%%

\section{RESULTS} \label{sec:results}
Figure \ref{fig:fig1}(a)
shows the C$^{18}$O intensity distribution of S235AB
integrated over the velocity range $-20<V_{\rm LSR}<-14$ km s$^{-1}$.
The clump has a well-defined single peak between the two small {H{$\,${\sc ii}}} regions (S235A and S235B),
and has an elliptical shape with an intense ridge elongating toward
the north-east direction.
Fainter outskirt extends to the south
toward another small {H{$\,${\sc ii}}} region named S235C
(see Figure \ref{fig:12co_dist}(d) in the Appendix \ref{sec:appendixa}).

We should note that the region shown in Figure \ref{fig:fig1}(a) has been observed
by \citet[][see their Figure 3]{Higuchi2009}. Their C$^{18}$O map appears slightly different from ours
showing a double-peaked structure, while our map in Figure \ref{fig:fig1}(a) shows a single peak
between S235A and S235B. We believe that this is simply because of the different velocity ranges
used for the integration. Though the velocity range used for the integration is not stated in their paper,
our map would appear very similar to theirs if we adopt a narrower
velocity range ($-18<V_{\rm LSR}<-15$ km s$^{-1}$).
\citet{Higuchi2009} argue that a clump-clump collision has induced cluster formation in the S235AB region,
and they classified the clump as `Type C' (i.e., the later stage of cluster formation) based on the morphology.
In the S235AB region, however, many young stellar objects (YSOs) have been discovered by \citet[][see their Figure 10]{Dewangan2011},
and their results show that the source S235AB-MIR is a very young massive protostar which is not yet able to excite an {H{$\,${\sc ii}}} region.
The source is apparently located at the center of the clump, indicating that the clump is
most likely to be the natal clump of the source. We therefore regard that the clump shown in Figure \ref{fig:fig1}(a)
is a distinct system not directly associated with the S235A or S235B {H{$\,${\sc ii}}} region, and that 
a new cluster-formation has just initiated in this clump, following cluster-formation in the two {H{$\,${\sc ii}}} regions.

In Figure \ref{fig:spectra}, we show an example of the $^{12}$CO, $^{13}$CO, and C$^{18}$O
spectra observed around the peak position of the clump.
While the optically thick $^{12}$CO spectrum
consists of two or more velocity components,
the optically much thinner C$^{18}$O and $^{13}$CO spectra
apparently have a single component at $V_{\rm LSR}\simeq -16.7$ km s$^{-1}$,
which should be the main component of the clump.
In the following analyses, we will focus on this component, and will briefly show
the other minor velocity components detected mainly in $^{12}$CO
in the Appendix \ref{sec:appendixa}.

To estimate the molecular mass of the clump,
we first estimated the excitation temperature of the C$^{18}$O molecules
from the peak brightness temperature of the $^{12}$CO emission line
at each position in the mapped region as displayed in Figure \ref{fig:fig1}(b), and
we calculated the C$^{18}$O column density $N$(C$^{18}$O)
in a standard way on the assumption of the Local Thermodynamic Equilibrium
\citep[e.g., see Section 3.2 of][]{Shimoikura2013}.
We then converted $N$(C$^{18}$O) to the column density of the hydrogen molecules $N$(H$_2$) 
using the empirical relation $N$(C$^{18}$O)/$N$(H$_2$)$=1.7\times 10^{-7}$
\citep{Frerking1982}.
Resulting $N$(H$_2$) map is shown in Figure \ref{fig:pv}(a).
We defined the surface area of the clump $S$ ($=0.35$ pc$^{2}$)
by the contour drawn at the half of the peak $N$(H$_2$) value ($1.4 \times 10^{23}$ H$_2$ cm$^{-2}$),
and we also defined the mean clump radius $R_0$($=0.34$ pc) as $R_0=\sqrt{S/\pi}$.
The clump mass $M$ contained in $S$ is estimated to be 680 $M_\sun$.
To measure the ellipticity of the clump $e_{\rm obs}$, 
we fitted the $N$(H$_2$) distribution with a 2 dimensional elliptical 
Gaussian function, for which we found $e_{\rm obs}=0.58$.
The 2 dimensional Gaussian fit infers that the total mass of the clump including periphery outside $S$
to be 1340 $M_\sun$.
We summarize these observed properties in Table \ref{tab:obs}, and will use them
to model the clump in Section \ref{sec:model}. 

Concerning the velocity field, the clump exhibits a typical C$^{18}$O line width of $\Delta V=1.8$ km s$^{-1}$ (FWHM).
Peak radial velocities of the emission line
stay around $V_{\rm LSR}=-16.7$ km s$^{-1}$ which should represent
the systemic velocity ($V_{\rm sys}$) of the clump.
It is notable that, with respect to the systemic velocity, the observed velocities
of the emission line exhibit an interesting symmetry around the intensity peak position. 
The feature can be easily recognized in the position-velocity (PV) diagram
shown in Figure \ref{fig:pv}(b) taken along the major axis of the clump.
The figure shows a clear velocity gradient around the center of the clump
with two well-defined peaks at the positions
$\pm 0.35$\arcmin ($\simeq 3.8\times 10^4$ AU)
and at velocities
$V_{\rm LSR} \simeq -17.2$ and $-16.2$ km s$^{-1}$
(separated by $\pm 0.5$ km s$^{-1}$ from $V_{\rm sys}$),
and they get close to $V_{\rm sys}$ at the outer sides of the clump.
Figure \ref{fig:pv}(c) displays another PV diagram measured along the
minor axis of the clump, and we can see a slight velocity gradient.

We should note that the features of the PV diagrams, especially the one seen
along the major axis, is very similar
to what have been found for infalling envelopes with rotation
around single protostars \citep[e.g.,][]{Ohashi1997,Momose1998,Bernard1999}.
This indicates that the internal motion of the cluster-forming clump
should follow a mechanism similar to that for the envelopes of
single protostars in spite of the large difference in size and mass.

%%%%%%%%%%%%%%%%%%%%%%%%%%%%%%%%%%%
%                                         Model                                                      %
%%%%%%%%%%%%%%%%%%%%%%%%%%%%%%%%%%%
\section{MODEL} \label{sec:model}

To understand the velocity distributions observed in C$^{18}$O, 
we made a simple model of infalling clump with rotation
with analogue to the model proposed by \citet{Ohashi1997}.
Their model is for a thin disk around a single young low-mass star, and
motion of the disk is assumed to be entirely dominated by the central stellar mass,
because the disk mass is much smaller.
They therefore assume that $V_{\rm inf}$ the infalling velocity and
$V_{\rm rot}$ the rotational velocity are $V_{\rm inf} \propto r^{-0.5}$ and
$V_{\rm rot} \propto r^{-1}$, respectively,
for conservations of the total energy and angular momentum.
They also assume that the density of the disk $\rho$
follows that for a free-falling disk ($\rho \propto r^{-1.5}$).
Based on the simple model, they successfully accounted for
the observed velocity field of a disk around IRAS 04368+2557 in LDN 1527.

In the case of the cluster-forming clump studied here,
the velocity field of the clump should not be as simple as for the case of the low-mass star,
because the mass of the system cannot be represented by a single particle
at the center of the clump but is distributed over a wide region in volume.
In addition, in the inner region of the clump, there must be forces of
clump-support such as the turbulence and magnetic
field which are difficult to quantify at the moment.
Moreover, the clump may not be treated as a disk,
because it has apparently a certain thickness.

In this study, we therefore assume an ellipsoidal clump
with an ellipticity of $e_0$ as illustrated in Figure \ref{fig:model}, 
and assume that $\rho$, $V_{\rm inf}$, and  $V_{\rm rot}$
of the clump can be approximated by the following equations;
%%Equation 1%%%%%%%%%%
\begin{equation}
\label{eq:density}
%\rho (r) = {\rho _0}{\left[ {1 + {{\left( {\frac{r}{{{R_{\rm d}}}}} \right)}^2}} \right]^{ - 0.75}} ~~,
\rho (r) = {\rho _0}{\left[ {1 + {{\left( {\frac{r}{{{R_{\rm d}}}}} \right)}^2}} \right]^{ -\frac{\alpha}{2}}} ~~,
\end{equation} 
%%%%%%%%%%%%%%%%% 
%%Equation 2%%%%%%%%%%
\begin{equation}
\label{eq:infall_velocity}
%{V_{\rm inf}}(r) = V_{\inf }^0\left( {\frac{r}{{{R_{\rm v}}}}} \right){\left[ {1 + {{\left( {\frac{r}{{{R_{\rm v}}}}} \right)}^2}} \right]^{ - 0.75}} ~~,
{V_{\rm inf}}(r) = V_{\inf }^0\left( {\frac{r}{{{R_{\rm v}}}}} \right){\left[ {1 + {{\left( {\frac{r}{{{R_{\rm v}}}}} \right)}^2}} \right]^{ - \frac{1+\beta}{2}}} ~~,
\end{equation} 
%%%%%%%%%%%%%%%%% 
and
%%Equation 3%%%%%%%%%%
\begin{equation}
\label{eq:rotation_velocity}
%{V_{\rm rot}}(R) = V_{\rm rot}^0\left( {\frac{R}{{{R_{\rm v}}}}} \right){\left[ {1 + {{\left( {\frac{R}{{{R_{\rm v}}}}} \right)}^2}} \right]^{ - 1}}
{V_{\rm rot}}(R) = V_{\rm rot}^0\left( {\frac{R}{{{R_{\rm v}}}}} \right){\left[ {1 + {{\left( {\frac{R}{{{R_{\rm v}}}}} \right)}^2}} \right]^{ - \frac{1+\gamma}{2}}}
\end{equation} 
%%%%%%%%%%%%%%%%% 
where
$\alpha$, $\beta$, $\gamma$,
$\rho_0$, $V_{\rm inf}^0$, $V_{\rm rot}^0$, 
$R_{\rm d}$, and $R_{\rm v}$
are constants.  $\rho_0$ can be expressed as $\rho_0=\mu m_{\rm H} n_0$
where $\mu$, $m_{\rm H}$, and $n_0$ are the mean molecular weight taken to be 2.4, proton mass,
and number density of hydrogen molecules (H$_2$).
$r$ and $R$ are the distance from the center of the clump and the axis of rotation
(the $Z$ axis in Figure \ref{fig:model}), respectively,
and they are expressed as 
%%Equation 4%%%%%%%%%%
\begin{equation}
\label{eq:radius1}
r = \sqrt {{x^2} + {y^2} + \left( {\frac{z}{{{e_0}}}}\right)^2} ~~,
%r = \sqrt {{x^2} + {y^2} + ( z/e_0)^2} ~~,
\end{equation} 
%%%%%%%%%%%%%%%%% 
and
%%Equation 5%%%%%%%%%%
\begin{equation}
\label{eq:radius2}
R = \sqrt {{x^2} + {y^2}} ~~.
\end{equation} 
%%%%%%%%%%%%%%%%% 
Note that Equations (\ref{eq:density}), (\ref{eq:infall_velocity}), and (\ref{eq:rotation_velocity})
are approximated by
$\rho \propto r^{-\alpha}$,
$V_{\rm inf} \propto r^{-\beta}$,
and
$V_{\rm rot} \propto r^{-\gamma}$
for large values of $r$ and $R$.
To make the equations equivalent to those assumed by
\citet{Ohashi1997}, we assume $\alpha=1.5$, $\beta=0.5$, and $\gamma=1$
in this paper.
%Note that Equations (\ref{eq:density})--(\ref{eq:rotation_velocity}) are
%designed to be equivalent to those assumed by \citet{Ohashi1997}
%for large values of $r$ and $R$.

In order to reproduce the observed C$^{18}$O spectra,
% by the model,
we set a cube of $256^3$ pixels with a pixel size
of $2000$ AU (corresponding to $\sim1/20$ of the angular resolution of the observations), 
and calculated $\rho$, $V_{\rm inf}$, and $V_{\rm rot}$ at each pixel 
according to the equations, and then integrated the number of H$_2$ molecules as a function of velocity
along the line-of-sight of the observers who view the clump at an angle of $\theta$
with respect to the rotation axis of the clump (see Figure \ref{fig:model}).
Resulting spectra are smoothed with a Gaussian beam with a width of 39600 AU (FWHM)
and are resampled onto the 13500 AU grid, corresponding to the same beam size (22\arcsec) and 
grid ($7\farcs5$) of the observations at the assumed distance (1.8 kpc).
In the calculations,
we made integrations along the line-of-sight up to $r=2R_0$
from the center of the clump where $R_0$ is the observed clump radius mentioned earlier
(0.34 pc $\simeq 6.8\times10^4$ AU), and we also
imposed a velocity dispersion of $\Delta V = 1.8$ km s$^{-1}$ (FWHM) to the gas
contained in each pixel.

In our model, there are seven parameters in total,
i.e., $e_0$, $\theta$, and the five parameters
($\rho_0$, $V_{\rm inf}^0$, $V_{\rm rot}^0$, $R_{\rm d}$, and $R_{\rm v}$)
in Equations (\ref{eq:density})--(\ref{eq:rotation_velocity}).
It is noteworthy that we can set strong restrictions to some of the parameters
from the observed data.
First, there is a certain relation between $e_0$
and $\theta$ to reproduce the observed ellipticity $e_{\rm obs}$($=0.58$).
Second, ${R_{\rm d}}$ in Equation ($\ref{eq:density}$) is rather independent on
the other parameters and can be decided by comparing directly
with the observed column density, for which we found $R_{\rm d}=4.7 \times 10^{4}$ AU.
Third, $\rho_0$ (and thus $n_0$) can be decided by comparing with the observed peak
$N$(H$_2$) value ($=1.4 \times 10^{23}$ cm$^{-2}$).
Finally, the quantity $V_{\rm rot}(R_1)/$sin$\theta$ gives an estimate for 
$V_{\rm rot}^0$ in Equation (\ref{eq:rotation_velocity}) where $V_{\rm rot}(R_1)$
is the observed rotation velocity at a large radius of $r=R_1$
on the outer edge of the clump.
As seen in Figure \ref{fig:pv}(b),
we found $V_{\rm rot}/{\rm sin} \theta \simeq 0.5$ km s$^{-1}$ for $R_1=6.5\times 10^4$ AU ($= 0.6$\arcmin).

Under these restrictions, we fitted the observed PV diagrams in Figure \ref{fig:pv}
by varying $e_0$, $V_{\rm inf}^0$, and $R_{\rm v}$
to find a set of the parameters minimizing $\chi^2$.
%In order to better fit the symmetric peaks seen in Figure  \ref{fig:pv}(b)
%which are important features indicating the infalling motion with rotation,
%we applied weights in the fitting proportional to the fourth power of the
%values at each pixel of the PV diagrams.
Parameters best fitting the data determined in this manner
are summarized
in Table \ref{tab:model} together with uncertainties at the 90$\%$ confidence levels.
%The best parameters give a reduced $\chi^2$ (i.e., divided by the degree of freedom)of  $\sim5$.
In Figure \ref{fig:pv}(e) and (f), we show the PV diagrams for the best model
that can be directly compared with the observed PV diagrams in Figure \ref{fig:pv}(b) and (c).
Spectra taken from the model and observations are also compared in Figure \ref{fig:pv}(g)--(i).
As can be seen in the figure, the model reproduces the observed PV diagrams well,
though the clump is not an ideal ellipsoid but has apparent distortion in some aspects.
It should be important to point out that
the observed two well-defined peaks seen in Figure \ref{fig:pv}(b) and a slight velocity gradient seen in Figure \ref{fig:pv}(c) can be reproduced only when we assume the infalling motion of the clump, and cannot be reproduced by any sets of parameters with $V_{\rm inf}=0$,
indicating that the clump is actually collapsing.

In addition to the analyses based on the C$^{18}$O data,
the infalling motion of the clump can be supported also by the asymmetric shape of
the optically thick $^{12}$CO emission line (Figure \ref{fig:spectra})
exhibiting higher blue-shifted component compared to the red-shifted component,
which is a characteristic feature of collapsing cores \citep[e.g.,][]{Zhou1993}.
The feature can be better recognized in PV diagrams of $^{12}$CO
shown in the upper panels of Figure \ref{fig:12co} in the Appendix \ref{sec:appendixb} measured along the same cuts as
in Figure \ref{fig:pv}. 
As detailed in the Appendix \ref{sec:appendixb}, we further attempted to reproduce
the PV diagrams of $^{12}$CO based on the model parameters summarized in Table \ref{tab:model}
assuming a flat $^{12}$CO fractional abundance as well as a 3D distribution of the excitation
temperature peaking at one of the small {H{$\,${\sc ii}}} regions (S235A).
Resulting PV diagrams are displayed in the lower panels of Figure \ref{fig:12co}.
We found that the observed higher blue-shifted $^{12}$CO emission both along the major and minor axes
(indicated by arrows in the figure) can be reproduced only when we assume
the infalling motion ($V_{\rm inf} \ne 0$),
strongly supporting our conclusion that the clump is collapsing.

We should note that the best values for the parameters $R_{\rm d}$ and $R_{\rm v}$

in Table \ref{tab:model}

do not match each other, and $R_{\rm d}$ $(\simeq4.7 \times 10^4$ AU$)$ is significantly
lager than $R_{\rm v}$ $(\simeq6.4 \times10^3$ AU$)$
by an order of magnitude.
% (see Table \ref{tab:model}).
The parameters represent the radii where the density and velocity laws start to change,
and the relation $R_{\rm d} \gg R_{\rm v}$ is puzzling because
%we would expect that
they should take the same value theoretically.
% as they are similar solutions.
The reason for our finding  $R_{\rm d} \gg R_{\rm v}$
% the significantly larger $R_{\rm d}$ than $R_{\rm v}$
can be naturally understood if C$^{18}$O is less abundant around the center of the clump
due to the destruction by the FUV radiation
from nearby massive stars
\citep[e.g.,][]{Shimajiri2014},
or due to the adsorption onto dust in dense regions \citep[e.g.,][]{Bergin2002}.
The former effect should be dominant in this case,
because, as seen in Figure \ref{fig:fig1}(b),
a large fraction of the clump is apparently heated by
at least one of the small H{$\,${\sc ii}} regions (S235A)
exhibiting an excitation temperature of $\gtrsim40$ K
which should be high enough for the molecules to evaporate from dust.
In addition, there are several intermediate/massive YSOs including S235AB-MIR forming around the clump center
which may be the sources of FUV \citep[see Figure 10 and Table C8 of ][]{Dewangan2011}.
In any case, 
note that Equation (\ref{eq:density}) with the inferred  $R_{\rm d}$ represents the density
distribution of the C$^{18}$O molecules, not necessarily the true total mass of the H$_2$ molecules.
If we assume that the true $R_{\rm d}$ (for H$_2$) is the same as
the derived $R_{\rm v}(\simeq6.4 \times 10^3$ AU$)$, the molecular density at the center of the clump
should be rescaled to $n_0 \simeq 2\times 10^6$ H$_2$ cm$^{-3}$
(instead of $1.1 \times 10^5$ cm$^{-3}$ in Table \ref{tab:model}).
We will use these corrected values of $R_{\rm d}$ and $n_0$ to estimate physical quantities
of the clump in Section \ref{sec:discussion}.

Finally, let us check the validity of our assuming the indices $\alpha=1.5$, $\beta=0.5$,
and $\gamma=1$ in Equations (\ref{eq:density})--(\ref{eq:rotation_velocity}).
We adopted these assumptions not only for the consistency with the original Ohashi's model, but also for
a technical reason to reduce number of the free parameters because we cannot determine
too many parameters within a reasonable amount of calculation time.
In order to check their validity, we varied the indices and the other parameters around
the best values listed in Table \ref{tab:model} to compare with the observed PV diagrams in Figure \ref{fig:pv},
and found that the indices minimizing
$\chi^2$ are $\alpha=1.5^{+0.6}_{-0.2}$, $\beta=0.5^{+0.1}_{-0.2}$, and $\gamma=1.0^{+0.2}_{-0.1}$
where the uncertainties represent the ranges for the 90 \% confidence level.
Though the uncertainties for $\alpha$ and $\beta$ are rather large,
the results infer that our assumptions for the indices are likely to be plausible.

%%%%%%%%%%%%%%%%%%%%%%%%%%%%%%%%%%%
%                                         Discussion                                                %
%%%%%%%%%%%%%%%%%%%%%%%%%%%%%%%%%%%
\section{DISCUSSION}\label{sec:discussion}

Though the angular resolution of our observations is rather limited,
the model parameters best fitting the observed data provide us with various
important implications on the structures of the clump. 
For example, Equation (\ref{eq:infall_velocity}) with
the value of $V_{\rm inf}^0$ (=1.3 km s$^{-1})$ should give us an estimate of the mass accretion rate 
$dM_{\rm in}/dt$, i.e.,  the mass infalling onto or passing through the clump surface per unit time.
For the iso-density surface $S_c(r)$ at the characteristic clump radius
$r=1.0\times10^{5}$ AU ($\sim1'$ at 1800 pc),
an approximation $dM_{\rm in}/dt \simeq \rho S_c V_{\rm inf}$ yields
$1.2\times10^{-3}$ $M_\sun$ yr$^{-1}$  if we assume $R_{\rm d}=R_{\rm v}=6.4 \times 10^3$ AU,
suggesting that it would take only $\sim1\times10^6$ yr
to gather the observed clump mass ($\sim1000$ $M_\sun$)
from more diffuse interstellar medium surrounding the clump.
%, which is shorter
%the age of the associated cluster \citep[$\sim3$ Myr,][]{Camargo2011}.
%In the inner region of the clump, $dM_{\rm in}/dt$ estimated in this manner
%stays around $\sim1\times 10^{-3}$ $M_\sun$ yr$^{-1}$
%over a wide range of $r$, and starts to become smaller around
%$r\simeq 1\times10^4$ AU,
%and then rapidly decreases to zero at $r=0$.
%The change of $dM_{\rm in}/dt$ as a function of $r$ implies that the entire clump is collapsing dynamically, and the infalling motion decelerates at the clump center to form a dense core with a size and mass of  $0.05-0.1$ pc and $2-10$ $M_\sun$ at the center of the clump.
%The mass infall rate remains rather flat over a wide region of the clump,
%but becomes smaller quickly toward the clump enter around the radius
%$R_{\rm d}$ where the velocity and density laws
%start to change.
In the inner region of the clump, $dM_{\rm in}/dt$ estimated in this manner
stays around $\sim1\times 10^{-3}$ $M_\sun$ yr$^{-1}$ over a wide range of $r$
down to $r=R_{\rm v}$.
Note that $R_{\rm v}$ is the radius where the velocity and density laws
start to change and it should be related to the size of the core formed at the center of the clump.
%but becomes smaller quickly toward the clump center around the radius
%$R_{\rm d}$ where the velocity and density laws start to change,
%and then decreases to zero at $r=0$.
The mass enclosed in the derived $R_{\rm v}=6.4 \times 10^3$ AU is $\sim 4$ $M_\sun$,
and we would expect that a core of this size and mass should be formed at the center of the clump.
It is noteworthy that such a dense core has actually been
discovered around the most massive star in S235AB located
at the center of the clump through molecular and millimeter continuum
observations by \citet[][see their Table 4]{Felli2004},
which may be the direct parent core of the $\sim11$ $M_\sun$ star found by \citet{Dewangan2011}.
Here, we should note that the core size and mass estimated by Felli et al.
vary rather largely taking values in the range $0.03 - 0.1$ pc and $2.3-31$ $M_\sun$, respectively,
depending on the tracers they used, which is probably due to
the different critical densities of the molecular emission lines as well as
due to the ambiguous conversion factors to the total hydrogen column densities.
Our estimate of the core mass is also rather ambiguous, and it can vary in the range
from $\sim 0$ $M_\sun$ to $\sim12$ $M_\sun$ for the 90\% confidence level
of $R_{\rm v}$ in Table \ref{tab:model} ($0-9.6\times10^3$ AU).
Taking into these ambiguities, we believe that the core found by Felli et al. should
correspond to the dense region around the clump center within $R_{\rm v}$.

%For example, $V_{\rm inf}^0$ and $n_0$ yield an estimate of a mass infall rate of $\sim1 \times10^{-3}$ $M_\sun$ yr$^{-1}$.
%The mass infall rate remains rather flat over a wide region of the clump,
%but becomes smaller quickly toward the clump enter around the radius
%$R_{\rm d}$ where the velocity and density laws start to change. 
%In fact, such a dense core has actually been discovered by \citet[][see their Figure 4]{Felli2004}
%through their interferometric continuum and molecular line observations, 
%which may be the direct parent core of the $\sim11$ $M_\sun$ star found by \citet{Dewangan2011}.

%この文を直す（いくら案）
%Formation of such a 0.1 pc-sized core at the center of cluster-forming clumps has been predicted by
%\citet{Wang2010} who performed numerical simulations assuming a massive clump having a mass of
%$1600$ $M_\sun$ similar to S235AB. 

Formation of a massive star at the center of cluster-forming clumps has been
studied theoretically by \citet{Wang2010} who performed numerical simulations assuming a massive
clump having a mass of $1600$ $M_\sun$, similar to S235AB.
They found that the most massive member of cluster forms at the clump center,
which may determine the final size of the cluster by blowing away the natal clump.
Interestingly, they also found that the mass of the massive star is controlled by the mass of the large-scale
natal clump rather than by the core just around the star, and the mass accretion rate onto the star
is regulated by the feedback of an outflow generated by the star.

We should note that \citet{Wang2010} defined the core simply as a 0.1 pc-sized sphere in diameter around
the central star having no distinguishable dynamical differences compared to the surrounding natal clump,
and in that sense, it is not precisely the same core we discuss here, which follows different
density and velocity profiles from the rest of the clump.
However,
their findings are important, because, if their conclusions are right, the small core found
around the central star merely represent the mass transiently trapped around
the central star without directly giving much influence to the stellar mass, and the final stellar mass
and therefore the final size of the cluster can be decided mostly by the size (or the infall rate) of the natal clump.
Note that, in the case of S235AB, our best estimate for the core mass is only $\sim4$ $M_\sun$,
while the observed stellar mass is $\sim11$ $M_\sun$. Though the uncertainty in our core mass estimate is large,
these values suggest that the hypothesis of \citet{Wang2010} can be plausible.
%our maximum estimate of the core mass is only $\sim12$ $M_\sun$
%or that inferred by \citet{Felli2004} is $\sim31$ $M_\sun$ at most,
%suggesting that hypothesis of \citet{Wang2010} could be plausible.
%Note that, in the case of S235AB, the inferred core mass is only $\sim 4$ $M_\sun$ while the observed
%stellar mass is $\sim11$ $M_\sun$, suggesting that hypothesis of \citet{Wang2010} can be plausible.

Beside the infalling motion of the clump, there must be ejection of mass by outflows which should be
playing an important role in the cluster-forming clump. They have two major effects in the cluster formation.
One is to inject the turbulence to the natal clump, which prevents the clump from collapsing or
decelerates the infall velocity, and the other is to release the angular momentum of the mass infalling onto the central star.

%\citet{Felli2004} reported that there are two molecular outflows in S235AB.

Based on the HCO$^{+}(J=1-0)$ interferometric observations, \citet{Felli2004} found
two candidates of molecular outflows at the center of the clump:
One is an outflow with intense blue and red lobes extending toward the NE--SW direction, 
and the other is a fainter outflow extending toward the NNW--SSE direction (see their Figure 5 and Table 5).
We call them the NE--SW outflow and the NNW--SSE outflow in this paper, respectively.
%The high velocity lobes of these outflows are roughly perpendicular to each other
%exhibiting an interesting X shape structure.
Compared with the shape of the clump,
the NE--SW outflow is quite puzzling because it is extending along the $x_{\rm O}$ axis in Figure \ref{fig:pv},
orthogonal to the rotation axis of the clump.
Though there may be some possible mechanisms (e.g., precession of the outflow) to account for the mismatch,
%we do not understand the geometrical relationship between the NE--SW outflow and the clump, and
we suspect that the NE--SW outflow is not a real outflow, and that its blue and red lobes are tracing something else by chance:
For example, the blue lobe can be due to blue-shifted high velocity gas blowing from the small H{$\,${\sc ii}} region S235A,
and the red lobe can be due to contamination by distinct velocity component(s) unrelated to the clump.
We present some observational data to support this possibility in the Appendix \ref{sec:appendixa}.

On the other hand, we believe that the other NNW-SSE outflow is a real outflow,
not only because it is known to coincide well with a jet traced by water maser spots \citep{Burns2015},
but also because it's blue and red lobes are elongating along the rotation axis of the clump (i.e., the $y_{\rm O}$ axis in Figure \ref{fig:pv}).
%The apparent extension of the blue and red lobes is parallel to the rotation axis of the clump,
%suggesting that the outflow originates from the central star formed by the contraction of the clump.
The mass ejection rate of the NNW-SSE outflow is estimated by \citet{Felli2004} to be
$2.5\times10^{-4}$ $M_\sun$ yr$^{-1}$ (see their Table 5), indicating that roughly $20-30$ $\%$
of the infalling mass derived from our model parameters $(\sim1 \times10^{-3}$ $M_\sun$ yr$^{-1})$
are fed back to the natal clump.
The $20-30$ $\%$ feedback of the infalling mass meets well with what we would expect
from numerical simulations \citep{Shu1988,Pelletier1992}.
%以下はカット?
If we include the NE--SW outflow, however, the total mass ejection rate by the two outflows
would amount to $\gtrsim 1.55 \times10^{-3}$ $M_\sun$ yr$^{-1}$ \citep{Felli2004} which is larger
than the infalling mass by $\gtrsim 55$  $\%$. In that case, the clump would be in the final stage of cluster formation
being dispersed by the internal outflows, but we suspect that the NE--SW outflow might not be a real outflow as we mentioned.

We should also note that the infall velocity with the derived parameter 
$V^0_{\rm inf}=1.3$ km s$^{-1}$ is significantly smaller than
the Virial velocity calculated as $V_{\rm vir}=\sqrt{GM/R_0} \simeq  4$ km s$^{-1}$,
strongly indicating the existence of clump-supporting forces.
The injection of turbulence by the feedback of the outflow is a possible source
for the deceleration of the infall velocity.
It is interesting to note that a density profile of gravitationally stable clumps and
free-falling clumps should follow a power law $\rho \propto r^{-\alpha} $ with $\alpha=2$  and $1.5$, respectively,
but both of the values are within the uncertainty of our value for $\alpha$ ($=1.5^{+0.6}_{-0.2}$),
%but our value for $\alpha$ ($=1.5^{+0.6}_{-0.2}$) is rather ambiguous covering these two values,
which may be due to the deceleration of the infalling motion by the clump-supporting forces
including the turbulence generated by the outflow.

The other expected role of the outflow to release of the angular momentum
would be probed quantitatively by comparing our model parameter
$V_{\rm rot}^0$ and the angular momentum of the outflow,
though the measurement of the outflow rotation is not easy
because it requires a very high spatial resolution.
We expect that such a measurement can be done by a large interferometer like ALMA
in near future.

%<モデルの欠点とインフォール速度が遅すぎ／速すぎ>
Here, we discuss the major incompleteness of our model.
First, we assume a fixed ellipticity $e_0$ for the density distribution. However, while we tried to fit the
observed data, we realized that the ellipticity changes along with the radius being
likely to be more flattened in the inner region, which is not taken into account in our model.
In addition, for simplicity, we assume that the infall velocities always point
toward the center of the clump, which is apparently inappropriate
especially around the center of the clump \citep[e.g., see Figure 2 of][]{Nakamura1995}.
Finally, other than the self-gravity, there must be
some important sources that should give significant influence on the internal motion of the clump,
such as the turbulence and magnetic field as well as the feedback from
stars forming in the clump, but none of these effects is not taken into account in our model,
though they may actually give a significant influence on the infalling velocity of the clump
as discussed in the above.

%<磁場>
Among the possible effects not considered in the present model,
the magnetic field is of our particular interest, not only because it should affect
the shape and velocity fields of the clump by preventing it from collapsing,
but also because it may control the star formation rate.
Because the clump is actually forming a cluster, we would expect that the clump mass should be larger than the 
critical mass that can be supported by the magnetic field
($M_{\rm cr}/M_\sun)=0.13\Phi/\sqrt{G}\simeq 2\times10^2(B/30 \mu {\rm G})(R/1 {\rm pc})^2$ \citep[e.g.,][]
{Shu1987,Nakano1985}.
In that case, strength of the magnetic field in the clump should be less than
B $\simeq900$ $\mu$G.
Although it is difficult to quantify the strength of the magnetic field at the density range of the clump
\citep[$10^4-10^{6}$ cm$^{-3}$, e.g.,][]{Crutcher2010},
measurement of the Zeeman splitting of some molecular lines such as CCS$(J_{N}=4_3-3_2)$ with a sensitive
receiver \citep[e.g.,][]{Nakamura2015} would provide us more precise picture of the cluster-forming clump.
 
%Because our on-going statistical study shows that such features in the PV diagrams
%are often seen among massive cluster-forming clumps (Shimoikura et al. 2016, inpreparation),
%we suggest that the dynamical infall with rotation is a common phenomenon
%in an early stage of cluster formation.

Finally, we should note that the characteristic feature in the PV map, i.e., the two-well defined
peaks as seen in Figure \ref{fig:pv}(b),
is often found in massive cluster-forming clumps (Shimoikura et al. 2016, in preparation).
We therefore suggest that the infalling motion with rotation 
is a common process in an early stage of cluster formation.

%%%%%%%%%%%%%%%%%%%%%%%%%%%%%%%%%%%
%                                         Conclusions                                              %
%%%%%%%%%%%%%%%%%%%%%%%%%%%%%%%%%%%

\section{CONCLUSIONS}

We observed the massive cluster-forming clump S235AB mainly in C$^{18}$O($J=1-0$) 
with the 45m telescope at the Nobeyama Radio Observatory. 
The observations have revealed that the clump has an elliptic shape having a mass of 
$\sim 1000$ $M_\sun$ and a radius of $\sim0.5$ pc.
The C$^{18}$O spectra of the clump often show double-peaks separated by $\sim 1$ km s$^{-1}$ 
and they are symmetrically located with respect to the center of the clump, 
which can be well recognized as the two well-defined peaks in the Position-Velocity (PV) diagram 
taken along the major axis of the clump. 
In addition, there is a slight velocity gradient along the minor axis of the clump. 
%The feature seen in the PV diagram is very similar to that found for an infalling core 
%with rotation around single low-mass YSOs. 
These features seen in the PV diagrams are very similar to those observed toward
small dense cores forming single low-mass YSOs, which can be well interpreted
as an infalling/collapsing motion of the cores with rotation. 

With analogue a model for single protostars proposed by \citet{Ohashi1997},
we made a simple model of the density and velocity distributions of the clump to fit the observed data.
The symmetric peaks in the PV diagram measured along the major axis as well as the velocity
gradient along the minor axis can be reproduced only when we assume an infall motion,
and cannot be reproduced by any sets of the parameters with the infalling velocity $V_{\rm inf}=0$,
indicating that the cluster-forming clump is actually collapsing toward the clump center. 
The mass infalling rate inside of the clump is large ($\sim 1 \times 10^{-3}$ $M_\sun$ yr$^{-1}$)
and remains rather flat over a large region, 
and it decreases around the center of the clump
at a radius of $\sim 6400$ AU, suggesting that a small core (with a mass and size of $\sim4$ $M_\sun$
and $\lesssim 0.1$ pc) and then a massive star should be formed.
Actually, there have been found such a core and a massive star $(\sim11$ $M_\sun)$
at the center of the clump by earlier observations.
All of these observational facts strongly imply that the clump S235AB should be
infalling with rotation following simple kinetic laws in the same way
as the cores around single YSOs.

%%%%%%%%%%%%%%%%%%%%
%             Acknowledgements               %
%%%%%%%%%%%%%%%%%%%%

\acknowledgments
We are very grateful to the anonymous referee for his/her helpful comments and suggestions to improve this paper. 
This research was financially supported by Grant-in-Aid for Scientific Research
(Nos.  24244017, 26287030, 26350186, 26400233, and 26610045).
The 45m radio telescope is operated by NRO, a branch of National Astronomical Observatory of Japan.

%\clearpage

%%%%%%%%%%%%%%%%%%%%%
%                           Tables                       %
%%%%%%%%%%%%%%%%%%%%%

% Table 1 (Observed)%%%%%%%%%%%%%%%%%%%%%%%%%%%%%%%%%%%%%%%%%%%%%%%%%%%%%%%
\clearpage

\begin{deluxetable*}{lcl} 
\tablecolumns{2} 
\tabletypesize{\scriptsize}
\tablecaption{Observed Properties of the Clump \label{tab:obs}} 
\tablewidth{0pt}  
\tablehead{ \colhead{Quantities} & \colhead{Values} & \colhead{Comments}
}
\startdata 
%\tableline
%\tableline
$S$ (pc$^2$)  &  0.35 & Surface area defined at the half of the peak $N$(H$_2$) value. \\
$R_0$ (pc)  &  0.34 & Mean radius calculated as $\sqrt{S/\pi}$.	\\
$M$	($M_{\sun}$)  & 680 & Mass contained in $S$. Total mass inferred from the Gaussian fit is 1340 $M_\sun$. \\
$N$(H$_2$)  (H$_2$ cm$^{-2}$) & $1.4 \times 10^{23}$ & Peak H$_2$ column density. \\
$V_{\rm sys}$ (km s$^{-1}$)  &  $-16.7$	& Systemic velocity. \\
$\Delta$V (km s$^{-1}$)  &  $1.8$ & Typical line width of the C$^{18}$O emission line (FWHM). \\
$e_{\rm obs}$ & $0.58$ & Apparent ellipticity.	\\
$V_{\rm rot}(R_1)/$sin$\theta$ (km s$^{-1}$) & $0.5$ & Apparent rotation velocity at $R_1=6.5 \times 10^4$ AU.	\\
\enddata 
\end{deluxetable*}

% Table 2 (Model)%%%%%%%%%%%%%%%%%%%%%%%%%%%%%%%%%%%%%%%%%%%%%%%%%%%%%%%
\clearpage

\begin{deluxetable*}{ll} 
\tablecolumns{2} 
\tabletypesize{\scriptsize}
\tablecaption{Model Parameters \label{tab:model}} 
\tablewidth{0pt}  
\tablehead{ 
\colhead{Parameters}   &  \colhead{Values (Range for the 90$\%$ confidence level)} 
}
\startdata 
$e_0$	&	$0.34$ ($0.27-0.41$) \\
$\theta$ (deg) & $65$ ($63-70$) \\
$n_0$ (H$_2$ cm$^{-3}$) &	$1.1 \times 10^5$ ($0.9\times10^5-1.3\times10^5$) $$ 	\\
$V_{\rm inf}^0$ (km s$^{-1}$)  & $1.3$ ($0.8-1.6$) \\
$V_{\rm rot}^0$ (km s$^{-1}$) & $5.4$ ($4.9-5.9$)	\\
$R_{\rm v}$ (AU) &	$6.4\times10^3$ ($0.0\times10^3-9.6\times10^3$) \\
${R_{\rm d}}$ (AU) & $4.7\times10^4$ ($3.2\times10^4-6.5\times10^4$) \\
\enddata 
\tablecomments{
The value of $n_0=1.1 \times 10^5$ H$_2$ cm$^{-3}$ should be rescaled to $\sim 2 \times 10^6$ H$_2$ cm$^{-3}$ for ${R_{\rm d}}=R_v=6.4\times10^3$ AU (see text).
}

\end{deluxetable*}

\clearpage

%%%%%%%%%%%%%%%%%%%%%
%                           Figures                      %
%%%%%%%%%%%%%%%%%%%%%

%%Fig 1%%%%%%%%%%%%%%%%%%%%%%%%%%%%%%%%%%

\begin{figure}
\begin{center}
\includegraphics[scale=1.0]{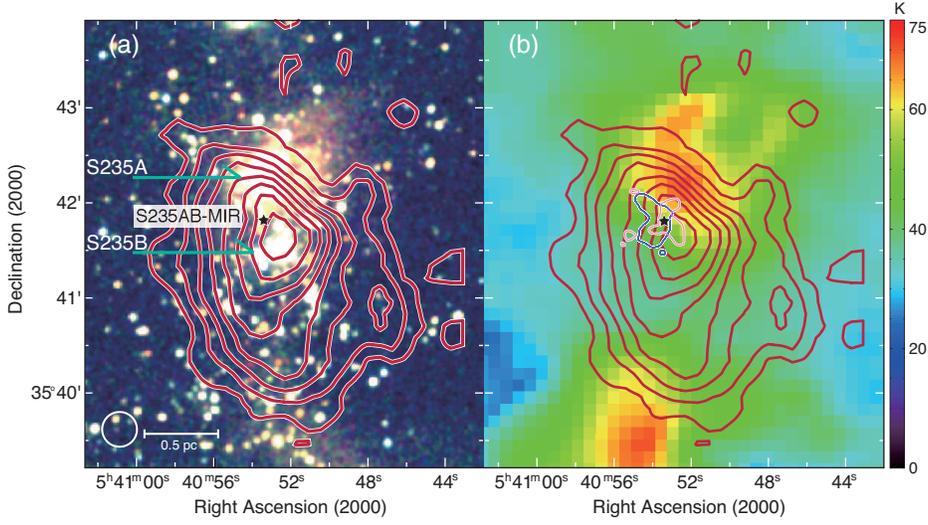}
\caption{
Distributions of the C$^{18}$O integrated intensity (contours) superposed on
(a) the NIR image taken from 2MASS and (b) the excitation temperature map
measured from the $^{12}$CO data. The velocity range used for the integration is
$-20<V_{\rm LSR}<-14$ km s$^{-1}$, and the lowest contours and contour intervals are
2.1 and 0.7 K km s$^{-1}$, respectively. In panel (a), two compact {H{$\,${\sc ii}}} regions
S235A and S235B are indicated by arrows, and  the position of S235AB-MIR the $\sim$11 $M_\sun$
star \citep{Dewangan2011} is shown by the star symbol. The open circle in the left-bottom corner
denotes the angular resolution of the C$^{18}$O map. In panel (b), outflow lobes reported
by \citet{Felli2004} are delineated by thin blue and pink lines (see their Figure 5).
\label{fig:fig1}}
\end{center}
\end{figure}
\clearpage

%%Fig 2%%%%%%%%%%%%%%%%%%%%%%%%%%%%%%%%%%

\begin{figure}
\begin{center}
\includegraphics[scale=1.0]{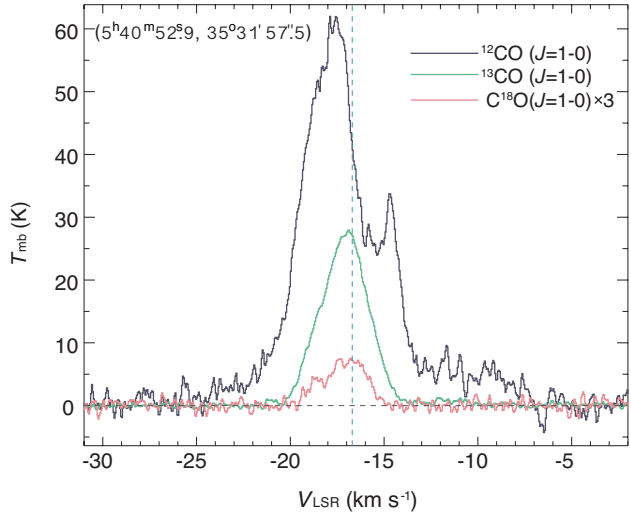}
\caption{
The $^{12}$CO, $^{13}$CO, and C$^{18}$O spectra observed toward
the peak position of the H$_2$ column density map shown in Figure \ref{fig:pv}(a).
Equatorial coordinates (J2000) of the peak position are given in the top-left corner.
The C$^{18}$O spectrum is scaled up by a factor of 3.
\label{fig:spectra}}
\end{center}
\end{figure}

%%Fig 3%%%%%%%%%%%%%%%%%%%%%%%%%%%%%%%%%%

\begin{figure*}
\begin{center}
\includegraphics[scale=1.0]{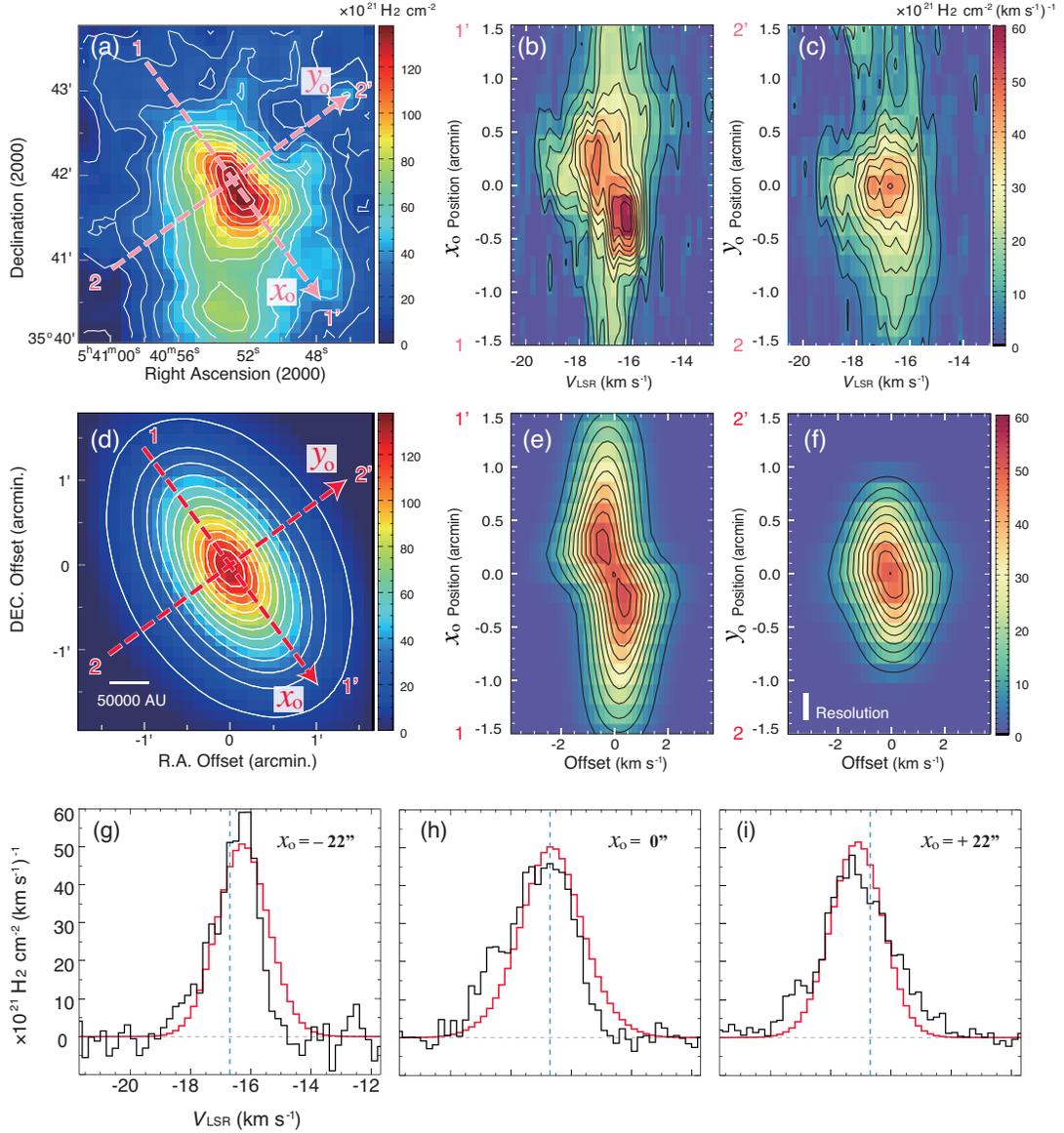}
\caption{
{\it Upper penels} : 
Distributions of the H$_2$ column density derived from the C$^{18}$O integrated intensity,
observed position-velocity diagram measured at the cut 1-1' along the $x_{\rm O}$ axis, and that measured
at the cut 2-2' along the $y_{\rm O}$ axis,
are displayed in panels (a)--(c).
The lowest contours and contour intervals are
$1 \times 10^{22}$ H$_2$ cm$^{-2}$ in panel (a),
and $5\times10^{21}$ H$_2$ cm$^{-2}$(km s$^{-1}$)$^{-1}$ in the other panels.
{\it Middle panels} :
Same as in the upper panels, but for the model best fitting the observed data.
Contours are the same as in the upper panels.
The center of the model clump is set to the peak position of the H$_2$ column density map in panel (a).
A linear scale for 50,000 AU is indicated in panel (d), and 
a common resolution for all of the position-velocity diagrams ($22''$ and $0.2$ km s$^{-1}$) is shown in panel (f).
{\it Lower panels} :
Observed (black lines) and model (red lines) spectra sampled at the center of the clump ($x_{\rm O}=0''$)
as well as at the positions $x_{\rm O}=\pm22''$ along the cut 1-1' corresponding
to the two peaks in panels (b) and (e).
The spectra are in units of the H$_2$ column density. The vertical broken lines denote the systemic
velocity ($V_{\rm sys}=-16.7$ km s$^{-1}$).
\label{fig:pv}}
\end{center}
\end{figure*}

\clearpage

%%Fig 4%%%%%%%%%%%%%%%%%%%%%%%%%%%%%%%%%%

\begin{figure}
\begin{center}
\includegraphics[scale=1.0]{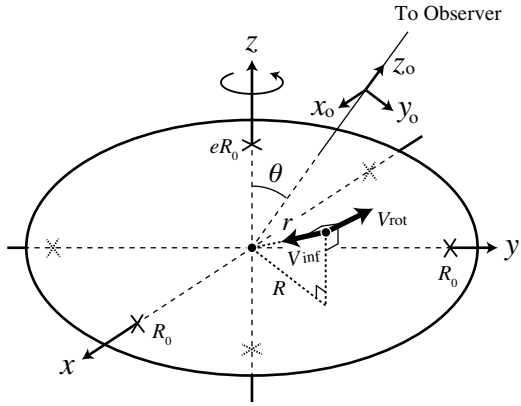}
\caption{
Schematic illustration of the model for cluster-forming clump.
We set the ($x,y,z$) coordinates taking the $z$ axis as the axis of rotation,
and set the other coordinates ($x_{\rm O},y_{\rm O},z_{\rm O}$)
rotated around the $x$ axis by $\theta$. The $z_{\rm O}$ axis points
toward the observer.
\label{fig:model}}
\end{center}
\end{figure}

\clearpage

%%Fig 5%%%%%%%%%%%%%%%%%%%%%%%%%%%%%%%%%%

\begin{figure*}
\begin{center}
\includegraphics[scale=0.8]{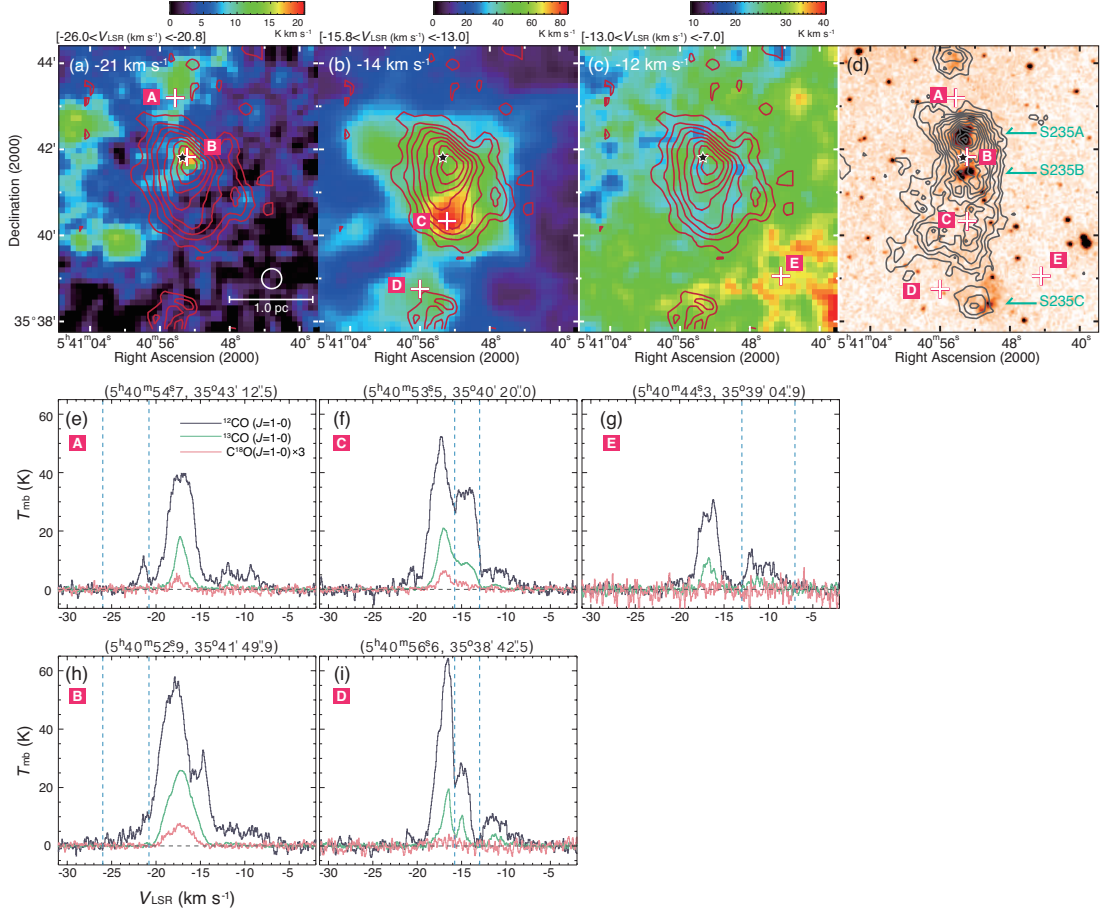}
\caption{
Panels (a)--(c) display the intensity maps of the $^{12}$CO emission line integrated over the velocity ranges
$-26.0<V_{\rm LSR}<-20.8$ km s$^{-1}$,
$-15.8<V_{\rm LSR}<-13.0$ km s$^{-1}$, and
$-13.0<V_{\rm LSR}<-7.0$ km s$^{-1}$ in this order.
The star symbol denotes the position of S235AB-MIR, and
the contours represent the C$^{18}$O integrated intensity same as in Figure \ref{fig:fig1}.
For comparison, we show in panel (d) the star density map produced from the 2MASS point sources
superposed on the $K_{\rm S}$ band image, which indicates the distributions of young stars forming in this region.
The lowest contours and contour intervals are 15 arcmin$^{-2}$ and 5 arcmin$^{-2}$, respectively.
Panels (e)--(i) show the $^{12}$CO, $^{13}$CO, and C$^{18}$O spectra
observed at the positions labeled A--E in panels (a)--(c). Equatorial coordinates (J2000) of the positions
are denoted at the top of the panels. The C$^{18}$O spectra
are scaled up by a factor of 3. The vertical broken lines denote the velocity ranges used for the integration
to produce the maps in panels (a)--(c).
\label{fig:12co_dist}}
\end{center}
\end{figure*}

\clearpage

%%Fig 6%%%%%%%%%%%%%%%%%%%%%%%%%%%%%%%%%%

\begin{figure}
\begin{center}
\includegraphics[scale=0.7]{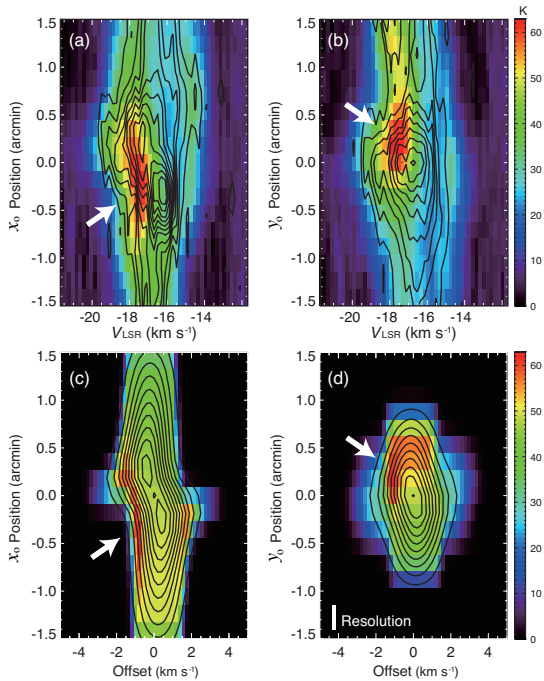}
\caption{
{\it Upper panels}: Observed position-velocity diagrams of the $^{12}$CO emission line (color scale)
measured along the cuts 1-1' and 2-2' in Figure \ref{fig:pv}.
{\it Lower panels}: Same as the upper panels, but for the simulated $^{12}$CO spectra (see the Appendix \ref{sec:appendixb}).
For comparison, contours for the position-velocity diagrams of $N$(H$_2$) displayed in Figure \ref{fig:pv}
are overlaid. White arrows indicate the blue-shifted $^{12}$CO emission exhibiting higher temperature
that can be reproduced only when we assume the infalling motion of the clump.
\label{fig:12co}}
\end{center}
\end{figure}

%%%%%%%%%%%%%%%%%%%%%
%                           Appendix                    %
%%%%%%%%%%%%%%%%%%%%%

\appendix

\section{Distributions of the Minor Velocity Components}\label{sec:appendixa}
In the observed $^{12}$CO, $^{13}$CO, and C$^{12}$O spectra,
the main velocity component tracing the massive clump studied in this work
is observed around the velocity $V_{\rm LSR}\simeq-16.7$ km s$^{-1}$.
In addition to this main component,
there are some other distinct velocity components in the observed region,
which we call `minor components' in this paper .
The minor components are fainter than the main component,
but they are significantly detected especially in the optically thick $^{12}$CO spectra.
In Figure \ref{fig:12co_dist}, we show the distributions of the minor components.
As seen in the figure, there are at least three minor components around the velocities
$V_{\rm LSR} \simeq -21$, $-14$, and $-12$ km s$^{-1}$. Their typical spectra are shown in
panels (e)--(i). In the following, we will describe these minor components to examine
their influence on our analyses in Section \ref{sec:model}.

The first minor component at $V_{\rm LSR} \simeq -21$ km s$^{-1}$ is widely seen mainly in the upper-left
side of panel (a) of Figure \ref{fig:12co_dist}, and its enhanced parts are distributed showing an arc-shaped
structure around the C$^{18}$O clump studied here (traced by the red contours).
It is unclear at the moment whether the minor component has a physical connection with the clump or not,
but it is likely to be associated with other larger clouds around the main Sh2-235 {H{$\,${\sc ii}}} region
located $\sim10\arcmin$ north to the clump, outside the maps in Figure \ref{fig:12co_dist} (Shimoikura et al. 2016, in preparation).
There is another peak at the position labeled B in panel (a) close to the center of the clump.
We display spectra observed at this position in panel (h).
The $^{12}$CO spectrum in the panel shows a wing-like feature over the velocity range indicated by the vertical broken lines in the panel.
This is likely to represent not only this minor component, but also the molecular outflows from the young stars forming there
as well as high velocity gas blowing from the small {H{$\,${\sc ii}}} regions S235A and/or S235B located at $\sim20\arcsec$ away as seen in panel (d).
Such a mixture of the outflowing gas from young stars and that blowing from small H{$\,${\sc ii}} regions
has actually been observed in $^{12}$CO in massive star forming regions \citep[e.g.,W40, see][]{Shimoikura2015},
and it is generally very difficult to distinguish them clearly in the observed spectra.
In the case of our data, a great fraction of the wing-like feature may be due to the high velocity gas blowing from
the two small {H{$\,${\sc ii}}} regions rather than the outflows from young stars, because its spatial distribution in panel (a)
tends to delineate the ridge or interface of the {H{$\,${\sc ii}}} regions.
As mentioned in Section \ref{sec:discussion}, the blue lobe of the NW-SE outflow \citep{Felli2004} seems to be
tracing the ridge of S235A (around the position B), and thus we wonder if the lobe is due to a real molecular outflow or the gas
blowing from the {H{$\,${\sc ii}}} region \citep[see also Figure 17 of][]{Felli2004}.

Distribution of the second minor component (at $V_{\rm LSR} \simeq -14$ km s$^{-1}$) in panel (b)
appears similar to that of the C$^{18}$O clump.
Though this is partially because the minor component and the main component (at $V_{\rm LSR}=-16.7$ km s$^{-1}$)
are close to each other in velocity and cannot be separated well
when generating the map in the panel, we believe that the minor component is also associated with the entire cloud
system in this region, because its spatial distribution traces well the distributions of the young stars
shown in the panel (d). The minor component is faint in the northern part of the observed region,
but it is prominent in the southern part, showing local peaks at the positions C and D in panel (b).
Velocities of the red lobe of the NW-SE outflow mentioned in Section \ref{sec:discussion} coincides with
this minor component in velocity, and we wonder if the red lobe might be due to a small clump of this component
\citep[see also Figure 4 of ][]{Felli2004}.

The third minor component  (at $V_{\rm LSR} \simeq -12$ km s$^{-1}$) is distributed all over the observed regions as
seen in panel (c), exhibiting a large velocity dispersion spreading over $-13 \lesssim V_{\rm LSR} \lesssim -7$ km s$^{-1}$.
This minor component may not be physically related to the clump studied here.

Among the three minor components, the second one at $V_{\rm LSR} \simeq -14$ km s$^{-1}$ can
affect our analyses in Section \ref{sec:model}, but its influence should be small because the component
is very week in C$^{18}$O over the main part of the clump. An enhancement of the C$^{18}$O emission
at $V_{\rm LSR}\simeq -14.5$ km s$^{-1}$ and $x_{\rm O}\simeq0.57\arcmin$ in Figure \ref{fig:pv}(b)
should be due to this minor component, but it doesn't cause much errors in our analyses as it is very faint.
The other two components at $V_{\rm LSR} \simeq -21$ and $-12$ km s$^{-1}$ should not affect our analyses either,
because they are well separated in velocity and are outside the velocity range in Figure \ref{fig:pv}(b) and (c)
used for the analyses. 
Though we cannot completely rule out the possible contamination
by the outflows and/or the high velocity gas blowing from the {H{$\,${\sc ii}} regions S235A and/or S235B
as seen at the position B in Figure \ref{fig:12co_dist}(a),
the C$^{18}$O emission line around the center of the clump is detected only in the velocity range
$-20\lesssim V_{\rm LSR}\lesssim-14.5$ km s$^{-1}$, and thus they are unlikely to give significant influence on our analyses.

\section{Position-Velocity Diagram of the $^{12}$CO emission line}\label{sec:appendixb}

Based on the model parameters in Table \ref{tab:model} derived from the C$^{18}$O data,
we further attempted to reproduce the observed PV diagrams of $^{12}$CO shown in the upper
panels of Figure \ref{fig:12co}. The diagrams are characterized by the systematically higher
blue-shifted components compared with the red-shifted components.
Our interest here is to investigate if we can reproduce this feature
with the model parameters derived from the C$^{18}$O data,
which will provide another support for the infalling motion of the rotating clump.

Unlike in the case of C$^{18}$O, it is not easy to model the PV diagrams of $^{12}$CO precisely,
because the line is very optically thick and the shape of the emission line could be easily affected
by the assumption on the 3D distribution of the excitation temperature $T_{\rm ex}$
which cannot be determined well.
Because the 2D distribution of $T_{\rm ex}$ peaks at one of the small {H{$\,${\sc ii}}}
regions S235A as seen in Figure \ref{fig:fig1}(b) and also because the {H{$\,${\sc ii}}} region appears  rather obscured on
the optical images such as DSS, we believe that S235A should be located rather in the back of the clump
and should dominate the overall distribution of $T_{\rm ex}$ around the clump. We therefore assume the
distribution of $T_{\rm ex}$ in 3D as
%%Equation B1%%%%%%%%%%
\begin{equation}
\label{eq:Tex}
{T_{\rm ex}} = \frac{{2{T_0}}}{{1 + \sqrt {r'/{R_{\rm T}}} }}
\end{equation} 
%%%%%%%%%%%%%%%%% 
where $T_0$ and $R_{\rm T}$ are constants, and $r'$ is the distance to the
center of the {H{$\,${\sc ii}}} region located at $(x,y,z)=(x_{\rm H},y_{\rm H},z_{\rm H})$
in the $(x,y,z)$ coordinates in Figure \ref{fig:model}.
The apparent center of the {H{$\,${\sc ii}}} region is
$(x_{\rm O},y_{\rm O}) \simeq (-25\arcsec,-12\arcsec)$
on the observer's axes, but its precise location along the line-of-sight (i.e., the $z_{\rm O}$ axis)
is unknown. We therefore tentatively assume
$(x_{\rm H},y_{\rm H},z_{\rm H}) \simeq(-25\arcsec,-15\arcsec,-20\arcsec) \simeq (-4.5\times10^4,-2.7\times10^4,-3.6\times10^4)$ AU
at the distance 1800 pc.
We also assume $T_0=120$ K and $R_{\rm T}=7.5\arcsec=1.35 \times 10^4$ AU which reproduce
the observed 2D distribution of $T_{\rm ex}$ in Figure \ref{fig:fig1}(b) well.

Using the clump parameters listed in Table \ref{tab:model}, we calculated
the expected $^{12}$CO spectra at each line-of-sight in the same way as we did for the C$^{18}$O data
in Section \ref{sec:model}, but by solving the radiative transfer
taking into account the optical depth estimated for a $^{12}$CO fractional abundance of
$1 \times 10^{-4}$ \citep{Frerking1982}, and resampled the spectra on the same grid
with the same angular resolution as those of the observations.

Resulting PV diagrams are compared with the observed ones in the Figure \ref{fig:12co}.
Though there are some arbitrary parameters such as the location of the {H{$\,${\sc ii}}} region ($z_{\rm H}$),
the observed higher blue-shifted components both along the major and minor axes of the clump
are reproduced well by the above calculations.
We should note that, like in the case of C$^{18}$O, the observed higher temperature in the blue-shifted
components in the PV diagrams of $^{12}$CO
can be reproduced only when we assume the infalling motion of the clump ($V_{\rm inf} \ne 0$).
It is also noteworthy that we calculated the PV diagrams for some different values of $T_0$, $R_{\rm T}$,
and $(x_{\rm H},y_{\rm H},z_{\rm H})$ to find that the important features in the PV diagrams
do not change qualitatively for small variations of these parameters.

In Figure \ref{fig:12co}, however, there are some noticeable differences between
the observed and simulated PV diagrams. The observed blue-shifted components along the major axis
in panel (a) (indicated by the arrow) are more widely distributed in velocity than the simulated ones in panel (c).
This is due to our imposing a constant velocity dispersion of $\Delta V=1.8$ km s$^{-1}$ in the same way
same as for the C$^{18}$O data in Section \ref{sec:model}. We would obtain a simulated PV diagram
more similar to the observed one if we impose a larger velocity dispersion, which should be taken
into account especially around the {H{$\,${\sc ii}}} region.
The simulated PV diagram in panel (d) measured along the minor axis exhibits rather high temperature
over a wider velocity range than the observed one at $y_0 \simeq 0.5\arcmin$.
This was caused by our integrating the $^{12}$CO emission
only up to the radius twice larger than the clump radius ($R_0$ in Table \ref{tab:obs}), neglecting the
diffuse gas outside $r=2R_0$ as mentioned in Section \ref{sec:model}.
Finally, in the velocity range $-15 \lesssim V_{\rm LSR} \lesssim -14$ km s$^{-1}$, there are bumps 
at the position $x_{\rm O} \simeq 0.7\arcmin$ in panel (a) and $y_{\rm O} \simeq 0.0\arcmin$ in panel (b).
The bumps are apparently due to the minor velocity component (at $-14$ km s$^{-1}$) mentioned in the Appendix \ref{sec:appendixa},
which is not taken into account in our calculations.

%%%%%%%%%%%%%%%%%%%%%
%                           References                   %
%%%%%%%%%%%%%%%%%%%%%

\clearpage

%\bibliography{S235}

\begin{thebibliography}{}
\expandafter\ifx\csname natexlab\endcsname\relax\def\natexlab#1{#1}\fi

\bibitem[{{Asayama} \& {Nakajima}(2013)}]{Asayama}
{Asayama}, S., \& {Nakajima}, T. 2013, \pasp, 125, 213

\bibitem[{{Barnes} {et~al.}(2010){Barnes}, {Yonekura}, {Ryder}, {Hopkins},
  {Miyamoto}, {Furukawa}, \& {Fukui}}]{Barnes2010}
{Barnes}, P.~J., {Yonekura}, Y., {Ryder}, S.~D., {et~al.} 2010, \mnras, 402, 73

\bibitem[{{Bergin} {et~al.}(2002){Bergin}, {Alves}, {Huard}, \&
  {Lada}}]{Bergin2002}
{Bergin}, E.~A., {Alves}, J., {Huard}, T., \& {Lada}, C.~J. 2002, \apjl, 570,
  L101

\bibitem[{{Bernard} {et~al.}(1999){Bernard}, {Dobashi}, \&
  {Momose}}]{Bernard1999}
{Bernard}, J.~P., {Dobashi}, K., \& {Momose}, M. 1999, \aap, 350, 197

\bibitem[{{Burns} {et~al.}(2015){Burns}, {Imai}, {Handa}, {Omodaka},
  {Nakagawa}, {Nagayama}, \& {Ueno}}]{Burns2015}
{Burns}, R.~A., {Imai}, H., {Handa}, T., {et~al.} 2015, \mnras, 453, 3163

\bibitem[{{Camargo} {et~al.}(2011){Camargo}, {Bonatto}, \&
  {Bica}}]{Camargo2011}
{Camargo}, D., {Bonatto}, C., \& {Bica}, E. 2011, \mnras, 416, 1522

\bibitem[{{Chavarr{\'{\i}}a} {et~al.}(2014){Chavarr{\'{\i}}a}, {Allen},
  {Brunt}, {Hora}, {Muench}, \& {Fazio}}]{Chavarra2014}
{Chavarr{\'{\i}}a}, L., {Allen}, L., {Brunt}, C., {et~al.} 2014, \mnras, 439,
  3719

\bibitem[{{Crutcher} {et~al.}(2010){Crutcher}, {Wandelt}, {Heiles},
  {Falgarone}, \& {Troland}}]{Crutcher2010}
{Crutcher}, R.~M., {Wandelt}, B., {Heiles}, C., {Falgarone}, E., \& {Troland},
  T.~H. 2010, \apj, 725, 466

\bibitem[{{Dewangan} \& {Anandarao}(2011)}]{Dewangan2011}
{Dewangan}, L.~K., \& {Anandarao}, B.~G. 2011, \mnras, 414, 1526

\bibitem[{{Dobashi}(2011)}]{Dobashi2011}
{Dobashi}, K. 2011, \pasj, 63, S1

\bibitem[{{Dobashi} {et~al.}(2013){Dobashi}, {Marshall}, {Shimoikura}, \&
  {Bernard}}]{Dobashi2013}
{Dobashi}, K., {Marshall}, D.~J., {Shimoikura}, T., \& {Bernard}, J.-P. 2013,
  \pasj, 65, doi:10.1093/pasj/65.2.31

\bibitem[{{Dobashi} {et~al.}(2014){Dobashi}, {Matsumoto}, {Shimoikura},
  {Saito}, {Akisato}, {Ohashi}, \& {Nakagomi}}]{Dobashi2014}
{Dobashi}, K., {Matsumoto}, T., {Shimoikura}, T., {et~al.} 2014, \apj, 797, 58

\bibitem[{{Dobashi} {et~al.}(2005){Dobashi}, {Uehara}, {Kandori}, {Sakurai},
  {Kaiden}, {Umemoto}, \& {Sato}}]{Dobashi2005}
{Dobashi}, K., {Uehara}, H., {Kandori}, R., {et~al.} 2005, \pasj, 57, S1

\bibitem[{{Evans} \& {Blair}(1981)}]{Evans1981}
{Evans}, II, N.~J., \& {Blair}, G.~N. 1981, \apj, 246, 394

\bibitem[{{Felli} {et~al.}(2004){Felli}, {Massi}, {Navarrini}, {Neri},
  {Cesaroni}, \& {Jenness}}]{Felli2004}
{Felli}, M., {Massi}, F., {Navarrini}, A., {et~al.} 2004, \aap, 420, 553

\bibitem[{{Felli} {et~al.}(2006){Felli}, {Massi}, {Robberto}, \& {Cesaroni}}]{Felli2006}
{Felli}, M., {Massi}, F., {Robberto}, M., {et~al.} 2006, \aap, 453, 911

\bibitem[{{Felli} {et~al.}(1997){Felli}, {Testi}, {Valdettaro}, \&
  {Wang}}]{Felli1997}
{Felli}, M., {Testi}, L., {Valdettaro}, R., \& {Wang}, J.-J. 1997, \aap, 320,
  594

\bibitem[{{Frerking} {et~al.}(1982){Frerking}, {Langer}, \&
  {Wilson}}]{Frerking1982}
{Frerking}, M.~A., {Langer}, W.~D., \& {Wilson}, R.~W. 1982, \apj, 262, 590

\bibitem[{{Higuchi} {et~al.}(2009){Higuchi}, {Kurono}, {Saito}, \&
  {Kawabe}}]{Higuchi2009}
{Higuchi}, A.~E., {Kurono}, Y., {Saito}, M., \& {Kawabe}, R. 2009, \apj, 705,
  468

\bibitem[{{Israel} \& {Felli}(1978)}]{Israel1978}
{Israel}, F.~P., \& {Felli}, M. 1978, \aap, 63, 325

\bibitem[{{Klein} {et~al.}(2005){Klein}, {Posselt}, {Schreyer}, {Forbrich}, \&
  {Henning}}]{Klein2005}
{Klein}, R., {Posselt}, B., {Schreyer}, K., {Forbrich}, J., \& {Henning}, T.
  2005, \apjs, 161, 361

\bibitem[{{Lada} {et~al.}(2003){Lada}, {Bergin}, {Alves}, \&
  {Huard}}]{Lada2003b}
{Lada}, C.~J., {Bergin}, E.~A., {Alves}, J.~F., \& {Huard}, T.~L. 2003, \apj,
  586, 286

\bibitem[{{Lada} \& {Lada}(2003)}]{Lada2003}
{Lada}, C.~J., \& {Lada}, E.~A. 2003, \araa, 41, 57

\bibitem[{{Matsumoto} {et~al.}(2015){Matsumoto}, {Dobashi}, \&
  {Shimoikura}}]{Matsumoto2015}
{Matsumoto}, T., {Dobashi}, K., \& {Shimoikura}, T. 2015, \apj, 801, 77

\bibitem[{{Momose} {et~al.}(1998){Momose}, {Ohashi}, {Kawabe}, {Nakano}, \&
  {Hayashi}}]{Momose1998}
{Momose}, M., {Ohashi}, N., {Kawabe}, R., {Nakano}, T., \& {Hayashi}, M. 1998,
  \apj, 504, 314

\bibitem[{{Nakamura} {et~al.}(1995){Nakamura}, {Hanawa}, \&
  {Nakano}}]{Nakamura1995}
{Nakamura}, F., {Hanawa}, T., \& {Nakano}, T. 1995, \apj, 444, 770

\bibitem[{{Nakamura} {et~al.}(2014){Nakamura}, {Sugitani}, {Tanaka},
  {Nishitani}, {Dobashi}, {Shimoikura}, {Shimajiri}, {Kawabe}, {Yonekura},
  {Mizuno}, {Kimura}, {Tokuda}, {Kozu}, {Okada}, {Hasegawa}, {Ogawa}, {Kameno},
  {Shinnaga}, {Momose}, {Nakajima}, {Onishi}, {Maezawa}, {Hirota}, {Takano},
  {Iono}, {Kuno}, \& {Yamamoto}}]{Nakamura2014}
{Nakamura}, F., {Sugitani}, K., {Tanaka}, T., {et~al.} 2014, \apjl, 791, L23

\bibitem[{{Nakamura} {et~al.}(2015){Nakamura}, {Ogawa}, {Yonekura}, {Kimura},
  {Okada}, {Kozu}, {Hasegawa}, {Tokuda}, {Ochiai}, {Mizuno}, {Dobashi},
  {Shimoikura}, {Kameno}, {Taniguchi}, {Shinnaga}, {Takano}, {Kawabe},
  {Nakajima}, {Iono}, {Kuno}, {Onishi}, {Momose}, \& {Yamamoto}}]{Nakamura2015}
{Nakamura}, F., {Ogawa}, H., {Yonekura}, Y., {et~al.} 2015, \pasj, 67, 117

\bibitem[{{Nakano} \& {Yoshida}(1986)}]{Nakano1986}
{Nakano}, M., \& {Yoshida}, S. 1986, \pasj, 38, 531

\bibitem[{{Nakano}(1985)}]{Nakano1985}
{Nakano}, T. 1985, \pasj, 37, 69

\bibitem[{{Ohashi} {et~al.}(1997){Ohashi}, {Hayashi}, {Ho}, \&
  {Momose}}]{Ohashi1997}
{Ohashi}, N., {Hayashi}, M., {Ho}, P.~T.~P., \& {Momose}, M. 1997, \apj, 475,
  211

\bibitem[{{Pelletier} \& {Pudritz}(1992)}]{Pelletier1992}
{Pelletier}, G., \& {Pudritz}, R.~E. 1992, \apj, 394, 117

\bibitem[{{Peretto} {et~al.}(2006){Peretto}, {Andr{\'e}}, \&
  {Belloche}}]{Peretto2006}
{Peretto}, N., {Andr{\'e}}, P., \& {Belloche}, A. 2006, \aap, 445, 979

\bibitem[{{Peretto} {et~al.}(2007){Peretto}, {Hennebelle}, \&
  {Andr{\'e}}}]{Peretto2007}
{Peretto}, N., {Hennebelle}, P., \& {Andr{\'e}}, P. 2007, \aap, 464, 983

\bibitem[{{Reiter} {et~al.}(2011){Reiter}, {Shirley}, {Wu}, {Brogan},
  {Wootten}, \& {Tatematsu}}]{Reiter2011}
{Reiter}, M., {Shirley}, Y.~L., {Wu}, J., {et~al.} 2011, \apj, 740, 40

\bibitem[{{Saito} {et~al.}(2007){Saito}, {Saito}, {Sunada}, \&
  {Yonekura}}]{saito2007}
{Saito}, H., {Saito}, M., {Sunada}, K., \& {Yonekura}, Y. 2007, \apj, 659, 459

\bibitem[{{Sawada} {et~al.}(2008){Sawada}, {Ikeda}, {Sunada}, {Kuno},
  {Kamazaki}, {Morita}, {Kurono}, {Koura}, {Abe}, {Kawase}, {Maekawa},
  {Horigome}, \& {Yanagisawa}}]{Sawada}
{Sawada}, T., {Ikeda}, N., {Sunada}, K., {et~al.} 2008, \pasj, 60, 445

\bibitem[{{Sharpless}(1959)}]{Sharpless}
{Sharpless}, S. 1959, \apjs, 4, 257

\bibitem[{{Shimajiri} {et~al.}(2014){Shimajiri}, {Kitamura}, {Saito}, {Momose},
  {Nakamura}, {Dobashi}, {Shimoikura}, {Nishitani}, {Yamabi}, {Hara},
  {Katakura}, {Tsukagoshi}, {Tanaka}, \& {Kawabe}}]{Shimajiri2014}
{Shimajiri}, Y., {Kitamura}, Y., {Saito}, M., {et~al.} 2014, \aap, 564, A68

\bibitem[{{Shimoikura} {et~al.}(2015){Shimoikura}, {Dobashi}, {Nakamura},
  {Hara}, {Tanaka}, {Shimajiri}, {Sugitani}, \& {Kawabe}}]{Shimoikura2015}
{Shimoikura}, T., {Dobashi}, K., {Nakamura}, F., {et~al.} 2015, \apj, 806, 201

\bibitem[{{Shimoikura} {et~al.}(2013){Shimoikura}, {Dobashi}, {Saito},
  {Matsumoto}, {Nakamura}, {Nishimura}, {Kimura}, {Onishi}, \&
  {Ogawa}}]{Shimoikura2013}
{Shimoikura}, T., {Dobashi}, K., {Saito}, H., {et~al.} 2013, \apj, 768, 72

\bibitem[{{Shu} {et~al.}(1987){Shu}, {Adams}, \& {Lizano}}]{Shu1987}
{Shu}, F.~H., {Adams}, F.~C., \& {Lizano}, S. 1987, \araa, 25, 23

\bibitem[{{Shu} {et~al.}(1988){Shu}, {Lizano}, {Ruden}, \& {Najita}}]{Shu1988}
{Shu}, F.~H., {Lizano}, S., {Ruden}, S.~P., \& {Najita}, J. 1988, \apjl, 328,
  L19

\bibitem[{{Smith} {et~al.}(2009){Smith}, {Longmore}, \& {Bonnell}}]{Smith2009}
{Smith}, R.~J., {Longmore}, S., \& {Bonnell}, I. 2009, \mnras, 400, 1775

\bibitem[{{Torii} {et~al.}(2011){Torii}, {Enokiya}, {Sano}, {Yoshiike},
  {Hanaoka}, {Ohama}, {Furukawa}, {Dawson}, {Moribe}, {Oishi}, {Nakashima},
  {Okuda}, {Yamamoto}, {Kawamura}, {Mizuno}, {Maezawa}, {Onishi}, {Mizuno}, \&
  {Fukui}}]{Torii2011}
{Torii}, K., {Enokiya}, R., {Sano}, H., {et~al.} 2011, \apj, 738, 46

\bibitem[{{Wang} {et~al.}(2010){Wang}, {Li}, {Abel}, \& {Nakamura}}]{Wang2010}
{Wang}, P., {Li}, Z.-Y., {Abel}, T., \& {Nakamura}, F. 2010, \apj, 709, 27

\bibitem[{{Zhou} {et~al.}(1993){Zhou}, {Evans}, {K\"ompe}, \& {Walmsley}}]{Zhou1993}
{Zhou}, S., {Evans II}, N. J., {K\"ompe}, C., \& {Walmsley}, C. M. 1993, \apj, 404, 232

\end{thebibliography}

\end{document}